# High Clockrate Free-space Optical In-Memory Computing

Yuanhao Liang[1,2], James Wang[1,2], Kaiwen Xue[1,2], Xinyi Ren[1,2], Ran Yin[1,2], Shaoyuan Ou[2], Lian Zhou[1,2], Yuan Li[2], Tobias Heuser[3], Niels Heermeier[3], Ian Christen[1], James A. Lott[3], Stephan Reitzenstein[3], Mengjie Yu[1,2], Zaijun Chen[1,2, *]

[1]Department of Electrical Engineering and Computing Sciences, University of California, Berkeley, CA 94720, United States
[2]Ming Hsieh Department of Electrical and Computer Engineering, University of Southern California, Los Angeles, CA 90089, United States
[3]Institut für Physik und Astronomie, Technische Universität Berlin, Berlin, Germany
[*]Author e-mail address: zaijun@berkeley.edu

## Abstract

The ability to process and act on data in real time is increasingly critical for applications ranging from autonomous vehicles, three-dimensional environmental sensing and remote robotics. However, the deployment of deep neural networks (DNNs) in edge devices is hindered by the lack of energy-efficient scalable computing hardware. Here, we introduce a fanout spatial time-of-flight optical neural network (FAST-ONN) that calculates billions of convolutions per second with ultralow latency and power consumption. This is enabled by the combination of high-speed dense arrays of vertical-cavity surface-emitting lasers (VCSELs) for input modulation with spatial light modulators of high pixel counts for in-memory weighting. In a three-dimensional optical system, parallel differential readout allows signed weight values accurate inference in a single shot. The performance is benchmarked with feature extraction in You-Only-Look-Once (YOLO) for convolution at 100 million frames per second (MFPS), and in-system backward propagation training with photonic reprogrammability. The VCSEL transmitters are implementable in any free-space optical computing systems to improve the clockrate to over gigahertz. The high scalability in device counts and channel parallelism enables a new avenue to scale up free space computing hardware.

## Introduction

The emergence of artificial intelligence models is revolutionizing information processing with high precision in perception, reasoning, and control [1]. Deploying these AI models to the edge devices enables smart sensors with information processed directly where data is generated, ensuring real-time responsiveness, reducing latency, and minimizing energy-intensive cloud-edge data transfers. However, edge devices operate in environments that are flooded with dynamic and complex information, such as three-dimensional (3D) sensing in autonomous driving [2,3], defense warfield [4,5], or remote robotics [6,7] (Fig. 1a). Processing this information can require powerful models with enormous size and complexity that are difficult to deploy on edge platforms, due to the limited computing power, compact form factors, and strict energy budgets [8,9], which are stretching the performance of existing processors. The heart of deep neural network (DNN) models lies in large-scale matrix-vector multiplication (MVM), which requires intensive data parallelism, where conventional central processing units suffer from the Von Neumann bottleneck, and state-of-the-art accelerated computing processors, including graphics processing units [10], tensor processing units [11], field-programmable gate arrays [12], and application-specific integrated circuits [13], allow to rely on the movement and storage of charges with electronic wires that lead to low clock rates and high thermal dissipation due to capacitive losses [14–16]. Scaling up the computing power with complementary metal-oxide semiconductor (CMOS) platforms further increases size-, weight- and power- (SWaP) constraints [17,18]. For example, many edge-deployed sensors – such as light- or microwave-based ranging devices and cameras on autonomous vehicles [19] – incur substantial power draw and thermal load merely to capture and pre-process raw data. This energy expenditure not only reduces operational range but also necessitates active cooling, compounding overall system power consumption and undermining reliability in mobile or remote deployments [20,21]. Therefore, performing high-speed, large-scale, and low latency matrix operations under strict SWaP constraints remains a key bottleneck for edge AI hardware.

Photonic solutions are emerging to accelerate computation of matrix algebra with advantages of ultrahigh clockrates, low-loss propagation and high parallelism. Recent progress on photonic integrated platforms [22–25] using cascaded Mach-Zehnder interferometers (MZIs [22–24]), wavelength multiplexed weight banks [26], parallel operations with phase charge materials [27,28] (PCMs) and thin-film lithium niobate photonics [29–31] have achieved high degrees of integration and programmability. Remarkable efforts have scaled up these systems to thousands of devices [32,33], but further scaling is limited due to the large device footprints, limited chip areas, fabrication errors, and control and packaging complexity. Alternatively, implementing free-space systems based on light propagation allows promising system scaling and massive parallelism within the abbreviation limit. Besides, weighting with 3D-printed diffractive elements [34,35] or spatial light modulators (SLMs) provide millions of individually controllable weights [36,37]. However, the computational clock rates relying on the input data transmitters, using 3D-printed images [38,39], digital micromirror devices (DMDs) [40,41], SLMs [42,43] and organic light-emitting diodes (OLEDs) [36,44] are typically slow (less than 100 kHz), and this has been improved in a recent breakthrough with micro light-emitting diodes (µLEDs) that improved the clock rate to the megahertz range [45]. However, the system throughput is limited by the low clockrates. To fully unlock the potential of light, here we propose FAST-ONN to speed up these systems, where it explores (1) densely-packed volume-manufactured VCSEL transmitter arrays for input activation; (2) spatial fanout coping with diffractive element for parallel processing; (3) large-scale programmable weighting for in-system training; (4) low latency image classification under the You-Only-Look-Once (YOLO) [46] algorithm to provide accurate feature extraction.

## Results

The FAST-ONN is efficient for general matrix multiplication (GEMM). FAST-ONN consists of an input encoding layer, a spatial fanout layer, a weighting layer, and a read-out photodetector (PD) array. Here, each subimage with $N$ pixels (Fig. 1d) $X_{(N×1)}$ is encoded to an array of $N$ VCSEL devices. The emitted optical signals from the VCSELs are replicated to $M$ copies through a diffractive optical element (DOE). Each multiplies a weight kernel consisting of $N$ pixels on a spatial light modulator. This element-wise multiplication is completed with light propagation throughput cascaded modulation and the accumulation occurs by summing the intensity of the $N$ beams onto a photodetector, which generates photocurrents $Y_M=I(m)\propto\sum_N W_{m,n}X_n$. At each clock cycle, $M$ high-speed detectors read out the matrix-vector products of $M$ kernels simultaneously. The signals captured by the detectors are digitized and processed through downstream neural layers and a classification head, which generates the recognition result for each subimage. In contrast to our previous approach of VCSELs in homodyne time-multiplexed coherent photoelectric multiplication [47], the FAST-ONN achieves multiply-accumulates with cascaded modulation that does not require precise coherent control or phase-amplitude coupling that limits the computing accuracy, while preserving the system simplicity and scalability.

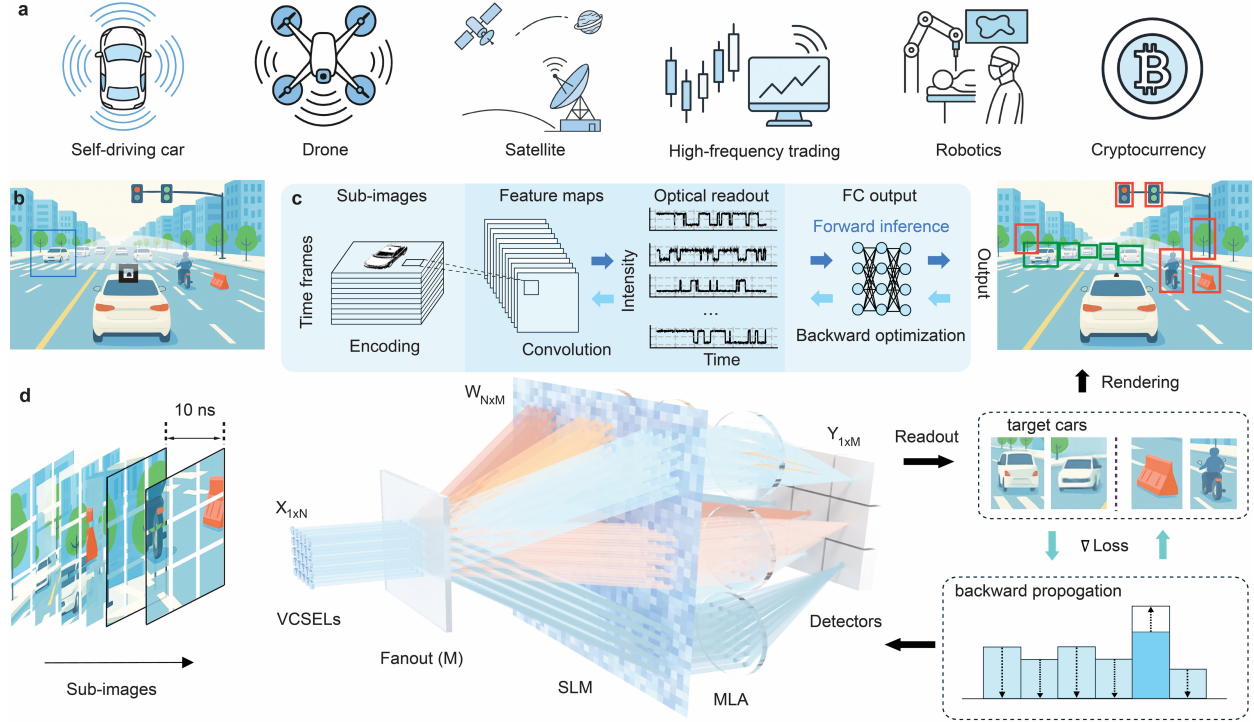

**Fig. 1. FAST-ONN concept. a**, Example edge-computing applications, including autonomous vehicles, drones, satellites, high-speed trading, robotics, and cryptocurrency. These scenarios span mobile, embedded, and high-frequency decision-making environments, where low-latency and energy-efficient AI inference is critical. **b**, Application scene. A car in the left-forward quadrant is detected and highlighted with a bounding box as the region of interest (ROI). This ROI is then routed into the dataflow for optical inference in (c). **c**, An edge camera catches images and processes them with parallel convolution to generate multiple feature maps, which yield multichannel optical readout signals feeding downstream outputs. Forward inference and backward optimization are both supported, enabling in-situ training. **d**, Sub-images are encoded onto a VCSEL array of N elements and fanned out into M parallel copies, where convolutional layers are executed on FAST-ONN; a trained model is loaded for inference, and the readout can be fed back to update weights for in-line parameter optimization. The resulting detections are rendered back onto the original scene, with cars outlined in green and other objects in red.

To benchmark its performance given the limited driver electronics, we implemented convolutional operations, which account for the vast majority of the total computational cost, often exceeding 70% in well-known algorithms like Visual Geometry Group (VGG) [48], ResNet [49], and YOLO [50]. In a convolutional neural network (CNN), kernels slide across the image, performing local multiplications with subimages to extract feature maps. Parallel operations in FAST-ONN allow simultaneous processing of multiple kernels at high clock rates. These computing results are rendered back to reconstruct the full scene, with task-specific labels (such as cars or non-car objects). Benefiting from the reconfigurability of the SLM, the system output can be utilized as feedback to update the spatial weighting, thereby enabling in-system parameter updates and real-time training tailored for different application scenarios such as self-driving cars, drones and robotics (Fig. 1 a-b).

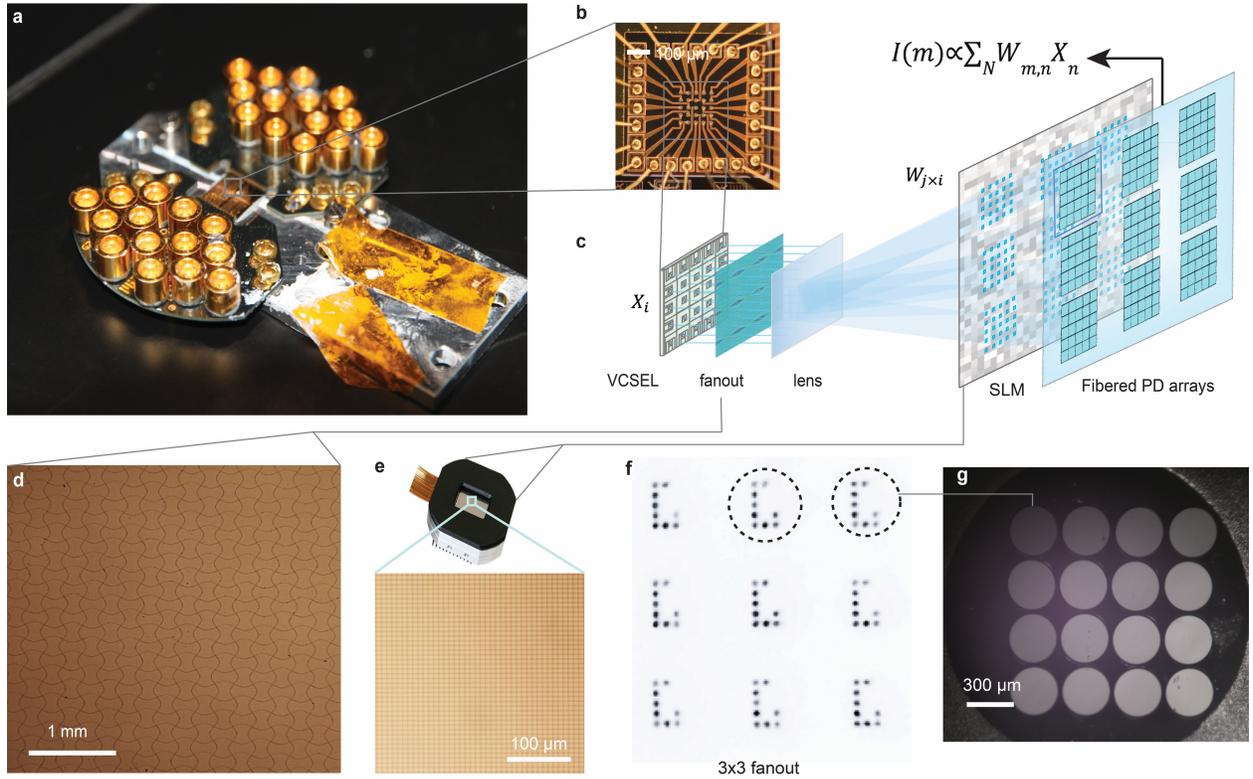

**Fig. 2. Experimental demonstration of FAST-ONN. a**, Fabricated arrays of VCSELs implemented in our FAST-ONN system with eight arrays each with 5 × 5 VCSELs. **b**, Zoomed-in view of an array of 5 × 5 VCSEL used in the experiment. **c**, Schematic of the FAST-ONN. The system comprises a VCSEL array that emits input-encoded optical signals, which are fanned out into multiple spatial copies via a DOE. These separated fanout beams are then modulated in parallel by a SLM to apply the desired convolution kernel weights. The weighted optical signals from each copy are summed by a multimode fiber with a core diameter of 300 μm and detected by a set of balanced photodetectors (BPD). **d**, Microscope image of the phase pattern encoded in the DOE used for 3 × 3 fanout. **e**, SLM and a zoomed-in view showing the individually-addressable liquid crystal pixels. **f**, Experimental 3 × 3 fan-out pattern of VCSEL array encoding a letter "C". **g**, Two-dimensional fiber array implemented in the system, one copy in the weighted arm and one in the reference arm.

Our FAST-ONN is developed with a VCSEL chip of eight arrays, each with 5 × 5 VCSELs (see Methods for details on the fabrication). The electrical contacts of the VCSELs are routed with short wires to maintain the bandwidth after wire bonding. In contrast to the linear array [45], our square array geometry supports both higher device counts and higher fanout factors at the aberration limit [43], as discussed in the Supplementary Information. Given the compact device size, a total of 400 VCSELs are fabricated in the chip area of about 10x10 mm² (Fig. 2a). The modulation bandwidth of the VCSELs is characterized to be 1 GHz (measured in [47]), limited in the present device design by the cavity lifetime. Noting that state-of-the-art VCSELs have achieved a bandwidth over 45 GHz [51]. Here, all the VCSELs are designed specifically with a slightly-elliptical cavity shape to emit linearly polarized light around 973 nm, which is important for efficient intensity modulation with the spatial light modulator (Methods). In our experiments, we use a single array with 25 VCSELs (Fig. 2b). While a single VCSEL can emit a maximum power of approximately 6 mW, we experimentally set them to about 2 mW per device to achieve our desired signal-to-noise ratio. The beams from the 5 × 5 array are fanned out to 3 × 3 copies using a diffractive optical element (Fig. 2d). For example, Fig. 2f shows the resulting beams when encoding a letter "C". Each copy with 5 × 5 beams is weighted on the spatial light modulator (Fig. 2e-f) and collected with a multimode fiber array (Fig. 2g) connected to an array of high-speed detectors.

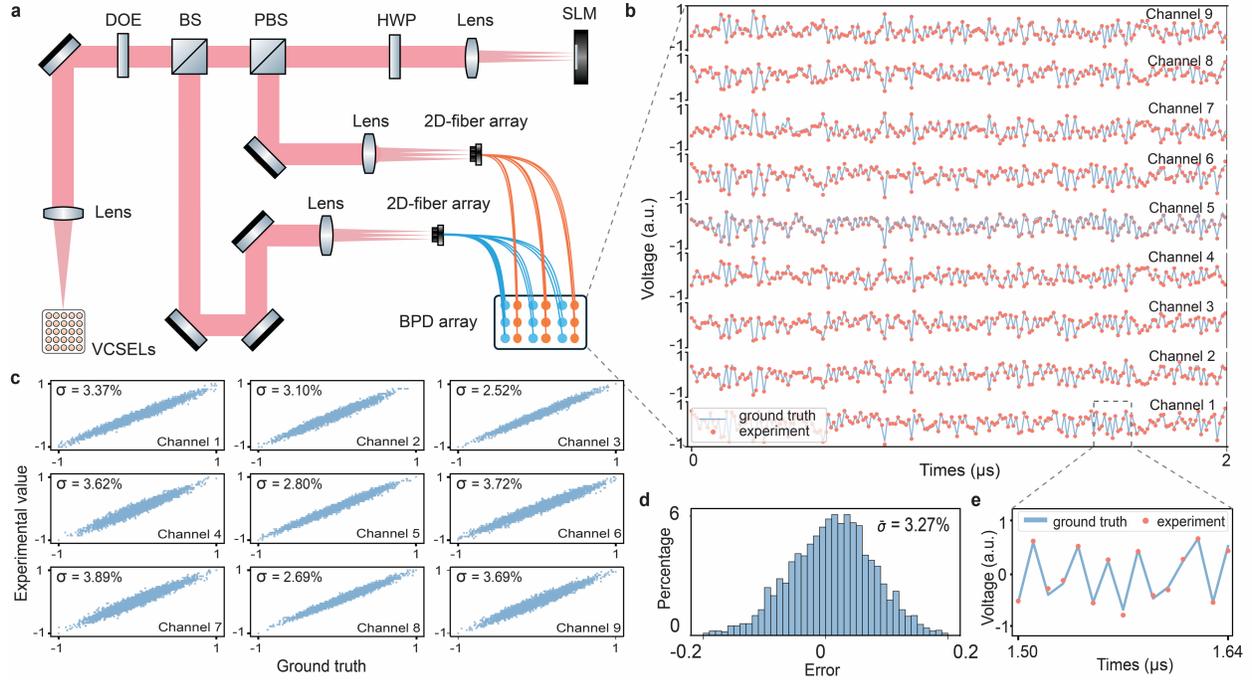

**Fig. 3. Simultaneous acquisition of random MVM with FAST-ONN. a**. FAST-ONN experimental setup. DOE: diffractive optical element; BS: beam splitter; PBS: polarized beam splitter; HWP: half waveplate; SLM: spatial light modulator; BPD: balanced photodetector. **b**, Comparison of experimental results and digital ground truth for the multiplication of a randomly distributed input vector with signed weights. The input is encoded onto a 9-VCSEL array operating at 100 MS/s, while the weights are encoded on the SLM. **c**, Correlation between ground truth and experimental outputs from the 9 optical channels. **d**, Histogram of multiplication errors across all optical channels, showing an average error of 3.27%. **e**, Enlarged view of experimental results versus ground truth. **f**, Edge detection results of the University of Southern California logo (from left to right: original logo; digital ground truth; FAST-ONN result). **g**, Edge detection results of the University of California, Berkeley logo (from left to right: original logo; digital ground truth; FAST-ONN result). **h**, Edge detection results on handwritten digits: the first row shows original digits; the second row shows digitally computed edges; the third row shows experimental results obtained by FAST-ONN.

A key challenge for intensity-based optical computing is encoding signed weight values, which are essential for neural network accuracy. FAST-ONN addresses this by using parallel reference beam and differential photodetection. As shown in Fig. 3a, each VCSEL beam is split into two paths: one directed to the SLM for polarization-based weight modulation (Methods), the other routed to generate 3 × 3 reference beams. Each SLM pixel is biased at the quadrature point, where equal signal and reference powers yield a weight of zero. Deviations (due to weight encoding) introduce power differences that map to positive or negative values in balanced detection. When parallelized, all 3 × 3 signal beams and their references are collected with multimode fibers (300 μm diameter) and fed to the positive and negative ports of 3 × 3 balanced photodetectors (BPD), followed by 9-channel digitization at 100 MS/s. This parallel differential detection extends FAST-ONN beyond unsigned intensity accumulation, enabling more general convolutional tasks.

## Parallel computing accuracy

We verify the computing accuracy in FAST-ONN by encoding a set of signed randomly distributed numbers at a rate of 100 MS/s to the array of nine VCSELs (N = 9). The 3 × 3 copies of the VCSEL beams are weighted with the SLM pixels of normal distributed randomly assigned values. The experimental time traces of duration of 2 μs, each with 200 MAC operations of all nine output channels, exhibited good agreement with numerical calculations (Fig. 3b to e). Analysis of the residuals $|y - ŷ|$ revealed an average computation error standard deviation of 3.27% across nine channels. We attribute the error to the crosstalk between the VCSELs beams on the SLM due to the limited numeral aperture (NA~0.064) of the coupling lenses, which can be improved with metalens arrays of high NA [52] in future system integration. This accuracy over 6 bits (5 bits for the numerical accuracy and 1 bit for the sign) is sufficient for most neural network tasks [53]. In image processing tasks, FAST-ONN allows real-time convolution of arbitrary input images. As an example, the university logos and handwritten digits are vectorized and encoded onto the VCSELs (Fig. 3f to h) at 100 MS/s. Applying an edge detection kernel on the SLM allows revealing the target information. The resulting edges are consistent with ground-truth with 95% accuracy.

## You only look once

Object-level car classification is a stringent, deployment-relevant benchmark for FAST-ONN. It matches YOLO-style tasks in autonomous driving and is central to intelligent transportations [54], where reliable car identification supports decision-making and safety [55]. Meeting these edge demands calls for fast, scalable, and robust semantics that strain conventional digital DNNs [56]. Here we benchmark FAST-ONN in accelerating YOLO tasks while maintaining accuracy. To test under realistic visual complexity, we adopt COCO dataset [57], as it provides diverse, fine-grained objects in cluttered scenes and is widely used for detection and classification in real-world settings.

We constructed an object-level car-vs-background classifier on COCO with ResNet-18 backbone to supply realistic inputs for evaluating FAST-ONN (Fig. 4a). Bounding-box-cropped patches are curated and resized under consistent quality and size constraints, and ResNet-18 is fine-tuned on COCO by unfreezing the last two residual stages and appending two lightweight convolutional layers to sharpen local features. The pipeline achieves accurate object-level classification, producing predictions with green bounding boxes indicating cars and red boxes indicating background (Fig. 4b), and attains an area under the ROC curve (AUC) of 0.98, demonstrating its effectiveness in handling challenging, cluttered real-world scenes. We implemented the convolutional layer of the pipeline on FAST-ONN. Using 2000 randomly selected test samples from COCO, we compared the outputs computed by the hybrid optical-electrical system with those from the purely electrical baseline. The two systems showed excellent agreement, with a standard deviation of 0.037 between outputs (Fig. 4c,d). Fig. 4e shows the probability distributions of car classification, which were obtained by feeding the outputs from the FAST-ONN and pure electrical inference into the convolutional layer of the system. Robustness under Gaussian noise was evaluated to simulate system performance under varying hardware imperfections (Methods). The performance gradually declines as the perturbation variance increases, yet remains stable with an AUC above 0.82 even at $σ = 0.5$, indicating a strong tolerance for hardware defects under the expected operating conditions.

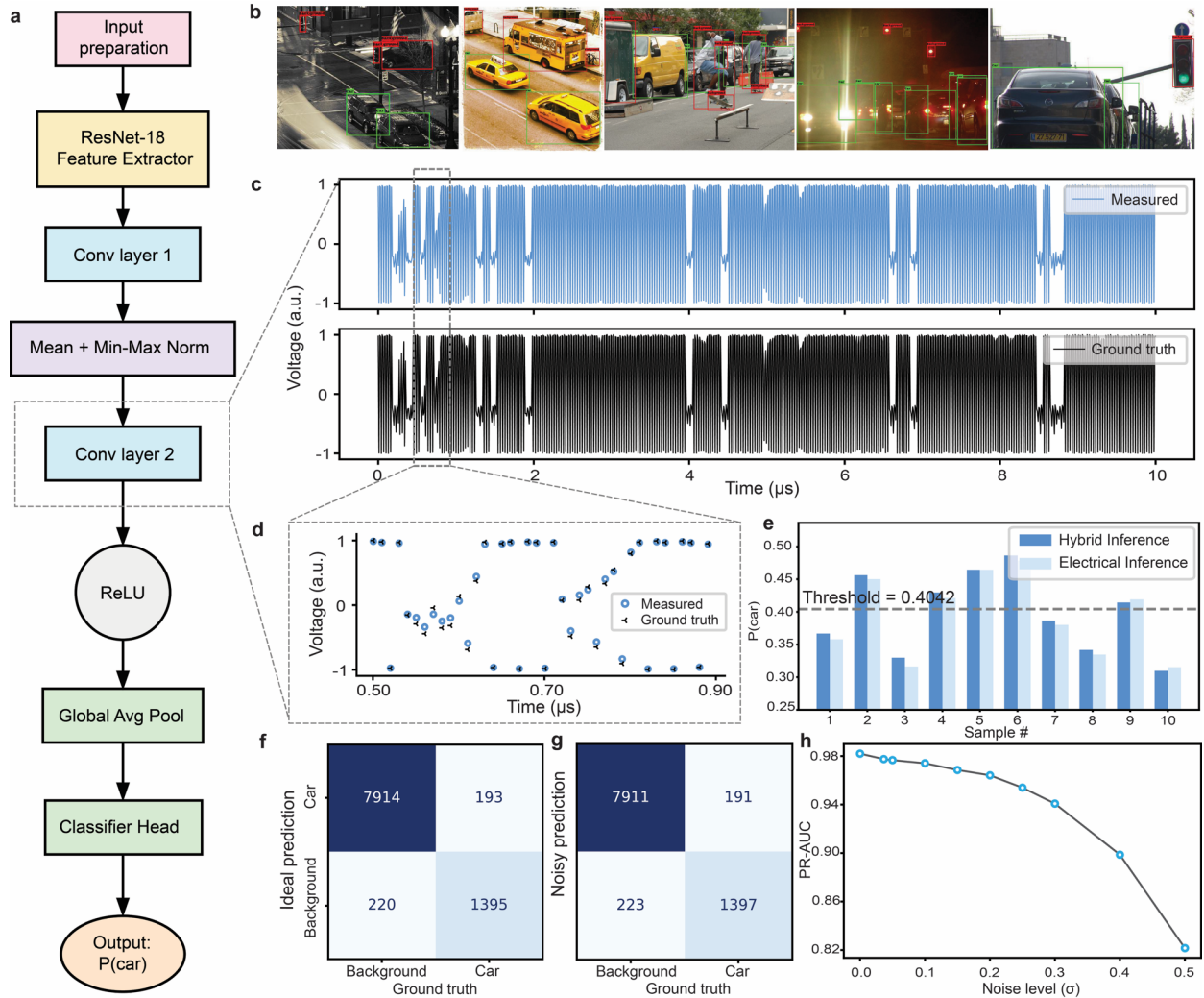

**Fig. 4. Object-level car classification and convolutional noise analysis using FAST-ONN.** The system evaluates whether a detected object corresponds to a car using a hybrid pipeline. **a**, Workflow of the object-level classification algorithm applied to the bounding-box-cropped images. **b**, Inference results with green boxes indicating objects identified as cars and red boxes denoting background. **c**, Time-trace of the layer-2 convolutional output from FAST-ONN compared with the electrical reference, with the error of 0.037. **d**, Zoom-in of a segment from (**c**), showing good consistency with the ground truth. **e**, Comparison of final classification probabilities with and without FAST-ONN integration, with a trained decision threshold of 0.4042. **f**, Confusion matrix of the full classification pipeline under ideal conditions. **g**, Confusion matrix obtained with experimental noise applied to all convolutional layers, including ResNet-18 layers 1-2 and Convolutional layers 1-2. **h**, AUC versus the injected noise level, ranging from $\sigma = 0$ to $\sigma = 0.5$.

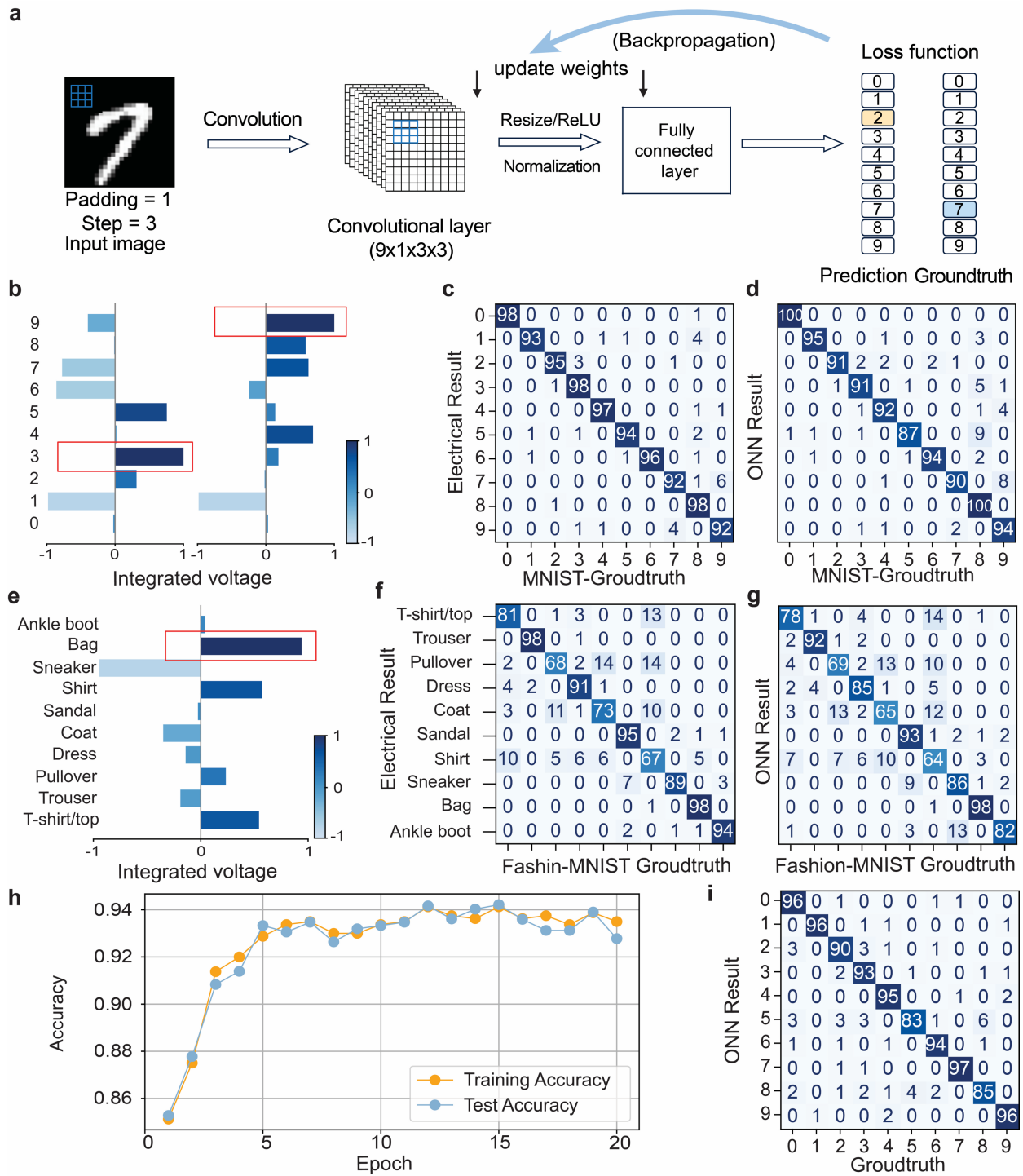

**Fig. 5. Benchmarking of machine learning inference and in-system training with FAST-ONN. a**, Neural network model for MNIST and Fashion-MNIST classification. The model consists of one convolutional layer with nine 3 × 3 kernels, followed by a fully connected output layer. Forward propagation results are compared against ground truth to compute the loss and update both the convolutional kernels and output weights. **b,e**, Output layer readout for MNIST (**b**) and Fashion-MNIST (**e**). The final classification is determined by selecting the label with the highest output value. **c,d**, Comparison of electrical (**c**) and optical (**d**) confusion matrices for MNIST classification with 800 randomly selected from test dataset (electrical 95.75%, optical 93.75%). **f,g** Comparison

of electrical (**f**) and optical (**g**) confusion matrices for Fashion-MNIST classification with 800 randomly selected from test dataset (electrical 84.88%, optical 80.75%). Optical and electronic results exhibit close agreement, validating the inference capability of FAST-ONN. **h**, In-line training accuracy versus epoch (20 epochs: 93.5% train subset, 92.8% test subset). **i**, Confusion matrix after 20 epochs of in-system training.

**CNN inference and training**

To benchmark its performance, FAST-ONN was evaluated on the standard 10-class Modified National Institute of Standards and Technology (MNIST) and Fashion-MNIST datasets. We train the digital reference on the full MNIST and Fashion-MNIST training sets with Adam and sparse categorical cross-entropy for 10 epochs, and evaluate on the held-out tests (Fig. 5b-g; training details are demonstrated in Methods). The digital model attains 95.75% on MNIST and 84.88% on Fashion-MNIST, while hardware inference achieves 93.75% and 80.75% respectively, closely matching the digital baseline. In addition to real-time inference, being able to adapt rapidly to the diverse environment changes is a necessity of edge intelligence, but it requires on-device, in-line training [23,58], to avoid long-haul data transfers and to reduce latency and energy overhead in scenarios such as autonomous driving cars, drones and real-time robotics. To realize this training in FAST-ONN, we couple optical forward passes at 100 MS/s with digital gradient updates, training on 800 randomly selected images from the MNIST training split (fixed seed) and evaluating on an independent 800-image subset from the test split. Inputs are forwarded through the optical convolution, cross-entropy loss is formed from digitized readouts, and weights are updated by gradient-based optimization under an element-wise [−1, 1] constraint. We computed the gradients digitally with ReLU gating and backpropagated through the sampled 3 × 3 patch structure while the forward path streams. After 20 epochs, the system reaches 93.5% accuracy on the training subset and 92.8% on the test subset (Fig. 5h–i), demonstrating practical on-device learning within the live inference path.

## Discussion

**Throughput**. We experimentally achieved optical convolution with a 5 × 5 VCSEL array in 3 × 3 parallel kernels at 100 mega subimages per second, corresponding to 0.9 billion convolutions per second in both training and inference tasks. Its performance is benchmarked with over 98% computing accuracy in a COCO classifier for inference of YOLO tasks and in-line training illustrates the hardware reprogrammability. The computational throughput in FAST-ONN $T = 2 \times N \times M \times R$ increases with VCSEL device counts ($N = 5 \times 5$) and fan-out factors ($M = 3 \times 3$). Currently the throughput $T = 45$ GOPS, limited by the low channel counts and clockspeed $R = 100$ MS/s. Based on the volume-manufactured device platforms and the large-scale spatial parallelism, a system with 32×32 VCSELs and 32×32 fanout factors may be achieved in the near term. As the detailed analysis in the Supplementary Fig. 2 discusses, these high channel counts are achievable with our compactly-arrayed VCSEL devices to enable large-scale fanout and make use of the high number (>1 million) of SLM weighting pixels available. With a VCSEL clock rate of 25 GS/s (VCSEL bandwidth over 45 GHz has been demonstrated [51]), a total throughput T>50,000 TOPS may become achievable, which would be over 10x the performance of an electronic system (4,000 TOPS in NVIDIA H100 [59]).

**Scalability**. A key requirement for the high fanout channels is the optical power for the readout signal-to-noise ratio to support the target computing bit precision. However, this does not set a constraint in our FAST-ONN because here the VCSELs serve as both laser sources and transmitters. Scaling up with high channel counts with VCSELs not only increases the data bandwidth, but also increases the total power. Our single-mode VCSELs can emit up to 6 mW, which results in a total power of over 6 W with the 32×32 VCSEL channels. The same power level would be challenging to reach with waveguide-based systems [60–63] due to power-dependent nonlinearities in the waveguides and on-chip lasers. As shown in Supplementary Fig. 3, we provide a detailed analysis regarding the computing bit precision, which suggests that 1 mW optical power on each detector is sufficient to support the signal-to-noise ratio of 7~8 bits computing at a clock rate of 25 GS/s. This only requires about 1 W total optical power for the 32×32 parallel channels.

Another key requirement is the O(N×M) channels of weighting devices, which is achievable with N × M = 1 million SLM pixels using off-the-shelf components (e.g., our SLM provides 1920 × 1200 = 2,304,000 pixels). Compared to on-chip integrated circuits, achieving millions of photonic devices (e.g., MZI mesh [22], memristive crossbar arrays [27,28], microring weighting banks [64]) is fundamentally challenging. Compared to other free-space systems that operate <10 MHz [34,36,39,42], FAST-ONN is based on a compact and high-speed VCSEL platform which supports 1000x improvement in clock speeds and dense arrays for high fanout copies.

**Energy efficiency**. We analyze the full-system energy consumption, including the electronic peripheral (Supplementary Fig. 1). Each *X*-vector from electronic memories is converted to drive the VCSELs using digital-to-analog converters (DACs). After optical fanout parallel computing, each MAC product is read out with a photodetector, followed by a transimpedance amplifier (TIA) and an analog-to-digital converter (ADC). The full-system electrical power, including the VCSEL drivers (bias power for laser generation and DAC drivers), the weighting power on the SLM pixels, and the readout electronics (Supplementary Table I). With appropriate electronic circuitry, our FAST-ONN system can achieve an energy efficiency of about 371 fJ/OP (~3 TOPS/W), which is similar to a state-of-the-art GPU. The energy cost for electronic-optical-electronic conversion can be further amortized by the high channel parallelism. If the system scales to the 32×32 VCSELs with 32×32 fanouts, an efficiency of 2 fJ/OP (500 TOP/W) might be reached (Supplementary Table I), which corresponds to 100x improvement compared to state-of-the-art electronic computing (~5 TOP/W in NVIDIA H100 [59]. Moreover, the matrix size over 1024x1024 can support fully connected layers beyond the convolutional operations.

In conclusion, FAST-ONN combines high-speed compact VCSEL transmitters with large-scale programmable spatial weighting, enabling reconfigurable and parallel AI computing. The compact, high-speed, fabrication-ready VCSEL platform can be exploited in any free-space optical computing system, including diffractive neural networks and optical generative models [34,65] and fanout systems [43,44] to improve the clock rate and thus throughput. FAST-ONN provides strong inference capabilities and supports backward optimization, enabling real-time tasks and has successfully implemented vehicle recognition in a hybrid optoelectronic pipeline. FAST-ONN is positioned to drive a compounding advance in capability and efficiency from edge deployments to large model regimes, charting a more efficient trajectory for computing.

## Methods

**Dataset.** The dataset of MNIST handwritten digits, MNIST fashion products, and COCO objects were respectively taken from http://yann.lecun.com/exdb/mnist/ and https://github.com/zalandoresearch/fashion-mnist, and https://cocodataset.org.

**VCSEL fabrication.** The VCSEL devices are based on semiconductor heterostructure microresonators, comprising a gain region formed by a stack of InGaAs quantum wells enclosed between two AlGaAs/GaAs distributed Bragg reflectors (DBRs) acting as cavity mirrors [66]. Each VCSEL within the 5 × 5 array was patterned via UV lithography and etched using an inductively coupled plasma reactive-ion process. To achieve high integration density, the array was arranged with a compact pitch of 80 μm between adjacent devices (Fig. 2a). Electrical interfacing was realized by depositing Au-based p-contacts on each cavity, which were individually wire-bonded to a printed circuit board connected to external driving electronics. All VCSELs in the array share a common ground connection to simplify electrical routing. A polymer cladding layer was applied over the chip to enhance mechanical stability, with selective openings above the apertures to allow light emission.

**Experimental setup.** The laser output from the VCSEL array is collimated by a converging lens (focal length = 150 mm) and collimated beam then passes through a DOE (MS-711-K-Y-X, Holo/OR Ltd) with a diffraction angle $\theta$ = 0.26 degrees. A beam splitter divides the beams to two beams for differential detection. One beam is focused onto the SLM (Santec Corp., Japan) by a lens with a focal length of 400 mm. The resulting optical spot formed on the SLM has a diameter of 19.7 μm, allowing it to be covered and modulated by a 3 × 3 array of SLM pixels. The center-to-center pitch between adjacent VCSELs, when projected onto the SLM is about 200 μm, and the fan-out center distance is around 1,800 μm. This ensures that the fan-out copies of the whole 5 × 5 VCSEL light spots are well-separated on the SLM (Fig. 2f). Following SLM modulation with a polarizing beam splitter, the optical beams are expanded to increase the center-to-center spacing of each fan-out copy to 330 μm in the image plane. This expansion enables one-to-one coupling between individual fan-out copies and the corresponding channels of a 2D fiber array (Beijing Reful Co. Ltd.), such that each fiber core receives a distinct spatial copy. An identical expansion is applied to the reference beam path, where a separate 2D fiber array captures the unweight replicas. Both fiber arrays have a common core pitch of 330 μm and a core diameter of 300 μm, ensuring efficient coupling with minimal crosstalk between adjacent channels. Each channel of the two 2D fiber arrays is routed to a balanced photodetector (Koheron, France), such that the two input ports of each BPD receive optical signals from corresponding channels in the modulated and reference paths, respectively. The electrical outputs from the BPD array are subsequently organized and routed through a patch panel, which facilitates stable cable management and flexible signal distribution before being connected to the acquisition card for digitization and processing.

**SLM calibration.** We pre-calibrate the PBS-SLM reflection so the reflected power is linear in the target weight $W_{n,m}$. For each pixel (or tile) we measure reflected intensity versus SLM gray level $g$, fit a monotonic curve, and invert it to a lookup table (LUT) that maps $W_{n,m}$ to the required $g$. During operation, the LUT enforces linearity. We apply a flat-field gain map to equalize the weighted power range and response across all fanout channels. At intervals, we sparsely re-measure a small subset of tiles and update the LUT and gains to compensate for slow drifts.

**Polarization modulation.** After calibration, we turn phase modulation into amplitude weights by polarization gating with a PBS-HWP-SLM stack (Fig. 3a). The VCSEL beams pass the PBS and a HWP rotates its linear polarization to 45 degrees relative to the SLM modulation axis. The SLM introduces a drive-dependent phase difference $\phi$ between eigenpolarizations; projection onto the PBS reflection port yields $I_R \propto \sin^2(\phi/2)$. Using the LUT $g(W_{n,m})$ and flat-field gains from SLM calibration, this response is linearized so the reflected power varies proportionally with $W_{n,m}$, and all fanout channels share the same linear response.

**Noise robustness.** To evaluate the tolerance for overall hardware imperfections in a portable way, we applied an additive zero-mean Gaussian perturbation (variance $\sigma^2$) after the activation of every convolutional layer (Resnet-18 backbone and the two task-specific convolutional layers). The evaluation was performed with 2000 test images to ensure statistical reliability. We used $\sigma = 0.037$ from Fig. 4g and scanned $\sigma$ in [0,0.5] in Fig. 4h to detect margin. This hardware aligned but device independent perturbation simulates the cumulative effects of PD/TIA/ADC readout noise, VCSEL power fluctuations, and SLM weight perturbations without relying on device specific calibration, enabling homogeneous comparisons between different model variants. Compared to the noise free baseline (Fig. 4f), the performance steadily decreased during the scanning process and maintained AUC above 0.82 at $\sigma = 0.5$ (Fig. 4h), demonstrating robustness under the intended operating conditions.

**CNN training.** We implement a compact convolutional classifier to match our intended hardware mapping. MNIST and Fashion-MNIST images (28 × 28 pixels) are normalized to [0, 1], reshaped to (28, 28, 1), and zero-padded by one pixel on all sides to form a 30 × 30 input. The network's core is a single bias-free 2D convolutional layer with nine 3 × 3 filters (stride = 3) followed by ReLU. This configuration produces nine 10 × 10 feature maps, which are flattened into a 900-dimensional vector and passed through a bias-free dense layer with softmax activation to yield ten class probabilities. Omitting biases streamlines our downstream optical implementation, and the chosen stride-and-padding scheme ensures that the network's spatial sampling aligns exactly with our FAST-ONN parallel convolution geometry.

After defining the architecture, we train the model using sparse categorical cross-entropy loss and the Adam [67] optimizer for 10 epochs on the full dataset, validating on the held-out test set at the end of each epoch. To respect hardware limits, we enforce an element-wise box constraint on all convolutional weights by clipping them to [–1, 1] after every gradient update. Once training converges, the nine learned 3 × 3 kernels are loaded onto the SLM as pixel-level weight masks.

## Acknowledgements
Z.C. acknowledges the DARPA NaPSAC program, and the National Science Foundation for the ASCENT program, the 2022 Sony Faculty Innovation Award, the 2025 Coherent II-VI foundation award, the 2023 Optical Foundation Challenge Award, and the Berkeley Baker fellows program.

## Author contributions

Z.C. conceived and supervised the project. Y.L., J.W. performed the experiments and finalized the experimental results, assisted by K.X., X.R., R.Y., L.Z., and Y. L.. T.H., N.H., J.A.L. and S.R. designed and fabricated the VCSEL arrays and characterized their performance. I.C. performed VCSEL packaging. Y.L. developed the software model for system-level neural networks. Y.L. and R.Y. carried out the theoretical analysis. J.W., L.Z., Y.L and M.Y. discussed and analyzed data. Z.C. and Y.L. prepared the manuscript. All authors contributed to the writing of the manuscript.

## Data availability

The datasets used and/or analyzed in the current study are available from the corresponding author upon reasonable request.

## Conflict of interest

Z.C and M.Y are cofounders of Opticore Inc. and hold equity. Other authors declare that they have no other competing interests.

# High Clockrate Free-space Optical In-Memory Computing: Supplementary Information

## System power consumption

We analyze the end-to-end full-system energy consumption, including both the optical power and the electronic interfaces, as summarized in Table I. A compute unit with the proposed electronic–optical interface is shown in Fig. S1. Digital activations are converted to analog by a high-speed digital-to-analog converter (DAC) that drives an input VCSEL; each VCSEL channel can be encoded with rates up to 25 GHz [1,2]. The emitted optical field is weighted by a liquid-crystal SLM pixel that implements the learned matrix element at that location. The weighted light is combined and detected by a photodiode, the photocurrent is amplified by a transimpedance amplifier (TIA), and the resulting voltage is digitized by an analog-to-digital converter (ADC) for readout.

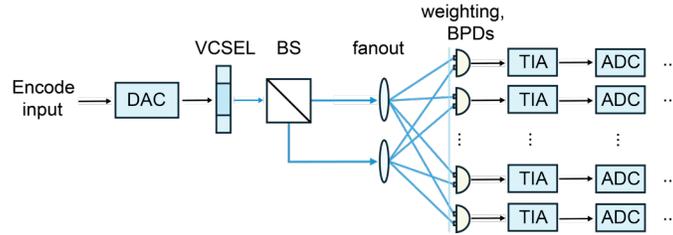

Supplementary Fig. 1. Proposed optoelectronic design. A Matrix-Vector multiplication unit.

For system-level estimates, we take $N$=32x32 = 1024 input VCSEL channels and M=32x32=1024 fan-out, each with its own photodetector, TIA, ADC, and nonlinear stage. This scale matches readily available SLM and source arrays and enables a fully parallel matrix-vector execution in a single cycle. The total cycle energy $E_{cycle}$ is the sum of $N$ DAC drives and VCSEL modulations (including wall-plug optical power), plus M TIA amplifications, ADC conversions, and nonlinear operations, together with one SLM refresh. Dividing $E_{cycle}$ by the N×M=$10^6$ MACs performed concurrently in that cycle gives the energy per operation. This 1/1000 (or $1/10^6$) amortization reduces picojoule-scale device energies to just a few femtojoules per MAC, yielding sub-picojoule operation energy even though individual component energies remain in the picojoule range. All numerical values used for this evaluation are listed in Table I.

Table I: Parameters in energy calculation

| Symbol | Parameter | Current system, R=100 MS/s, N=M=9 | | | Near-term development, R=25 GS/s, N=M=1000 | | |
|---|---|---|---|---|---|---|---|
| | | Energy cost | Parallelism | Energy efficiency | Energy cost | Parallelism | Energy efficiency |
| $E_{laser}$* | Electrical power for VCSEL generation | 400 µW | 2xRxM | 220 fJ/OP | 5 mW | 2xRxM | 0.1 fJ/OP |
| $E_{DAC}$ | Energy per DAC conversion | 0.5 pJ/conv [3] | 2xN | 25 fJ/OP | 0.5 pJ/conv [3] | 2xN | 0.25 fJ/OP |
| $E_{SLM}$† | Energy per pixel | 3 µW | 2xR | 15 fJ/OP | 3 µW | 2xR | 0.06 fJ/OP |
| $E_{TIA}$ | Energy per TIA | 180 fJ/conv. [4] | 2xM | 10 fJ/OP | 170 fJ/conv. [5] | 2xM | 0.085 fJ/OP |
| $E_{ADC}$ | Energy per ADC | 0.8 pJ/conv. [6] | 2xM | 44 fJ/OP | 2 pJ/conv. [7] | 2xM | 1 fJ/OP |
| $E_{NL}$ | Energy per nonlinearity | 1 pJ /OP [8] | 2xM | 55 fJ/OP | 1 pJ /OP [8] | 2xM | 0.5 fJ/OP |
| $E_{cycle}$†† | Total | – | – | 371 fJ/OP | | | 2 fJ/OP |

\* The VCSELs in the current system are biased around the threshold, with 1.3 V voltage and 300 µA current, resulting in an electrical power of about 400 µW. In the near-term development for high-speed readout, the VCSEL output power is expected to increase to above 1 mW. With a wall plug efficiency of 25%, we anticipate a power budget of 5 mW per VCSEL device.

† The SLM consumes about 8 W (for external supply, the controller electronics, and the LCOS panel) for a total of 1920x1200 pixels, resulting in a power cost of about 3 µW per pixel.

†† To obtain $E_{cycle}$ we divide the sustained operation rate by the energy per operation. Our system's throughput is Throughput = 2 × N × M × R, where N and M are the numbers of input VCSELs and weighting pixels, and R is the clock rate (assumed 25 GHz). Dividing this by the computed energy/OP yields the overall OP/s/W efficiency. The factor of 2 accounts for simultaneous multiplication and an accumulation.

**COCO car/background classifier with optical layer integration**

We derive a binary car/background set from COCO [9] to emulate realistic scene statistics while providing clean supervision for inserting an optical convolution. We crop positives around annotated car instances and discard boxes that are too small or have extreme aspect ratios; after clipping to image bounds, at least 70% of each box must be preserved. For negatives, we sample boxes from non-car categories within the same images. All crops are resized to 128×128 (RGB). Training uses RandomResizedCrop (scale 0.7–1.0, aspect ratio 0.75–1.33) and horizontal flipping, while validation employs a deterministic resize. Inputs are normalized with ImageNet statistics (mean = [0.485, 0.456, 0.406], std = [0.229, 0.224, 0.225]); data are shuffled and split 80/20 into train/validation.

The model builds on a ResNet-18 [10] pretrained on ImageNet. The stem and layer1 - layer2 remain frozen, and we fine-tune layer3 and layer4 together with a lightweight head that applies a convolution ("Conv layer1"; stride 1, padding 1, ReLU) to the feature map, followed by channel averaging and a per-sample min–max normalization to [0, 1] to align dynamic range with the optical front-end. Downstream, a compact bias-free convolution ("conv layer2") feeds a ReLU, global average pooling, a linear layer producing a scalar logit, and a learned scalar multiplier (logit_scale) for score calibration. This division leaves high-capacity feature extraction electronics while isolating a small spatial operator that can be executed optically without changing the task definition or evaluation protocol.

Training uses a focal binary cross-entropy objective ($\gamma = 2$) to accommodate class imbalance from hard negatives. We optimize with AdamW [11], mixed-precision, and parameter-grouped learning rates. Mini batches are 256 for training and 128 for validation; we train for 20 epochs with constant learning rates and pin_memory=True. Model selection is based on validation PR-AUC; the best checkpoint is retained, and the deployment threshold for the sigmoid score is chosen by maximizing the validation F1 along the precision–recall curve, yielding a reproducible operating point.

For optical deployment, the learned convolutional weights are programmed onto the SLM as pixel-level weight masks. During inference, the normalized feature map produced by the electronic backbone is streamed across the modulator in a sliding-window fashion: each receptive field is overlaid on the corresponding SLM cells, the SLM encodes the convolutional weights, and the modulated optical field is pipelined into the VCSEL transmitter and photodetection stack, thereby realizing the convolution optically while the remaining layers execute electronically for stability and compatibility with standard metrics. This hybrid path demonstrates that our system can execute learned spatial filtering as a real-time, pipelined optical primitive, supporting practical deployment in edge-computing scenarios where low latency and tight power budgets are paramount [12–14].

## System scalability

FAST-ONN offers strong potential for scalability: parallelism can be increased by expanding the number of VCSEL channels (N) and the spatial fanout factor (M), while overall throughput can be further enhanced by raising the VCSEL clock rate (R). A 32 × 32 VCSEL with a 32 × 32 fanout system is proposed in Fig. S2. In practice, scaling is constrained by two system-level considerations. The first is the power budget per copy, which is set by the pre-fanout optical power, transmission efficiency, and power for efficient readout. The second is aberration control across the wide field of view, since a grid with 1 degree spacing per order spans ±16 degrees, where off-axis aberrations must be carefully managed to preserve uniform weighting and avoid crosstalk.

**Power budget.** We target 10 µW per fanout copy at the weighing/readout plane. With 32 × 32 = 1024 fanout replica, the required optical power is 10 mW. This level is readily supported by microlens-collimated VCSEL arrays (one microlens per VCSEL for low divergence, high brightness) and multi-spot diffractive beam splitters that preserve beam quality across the lattice [15].

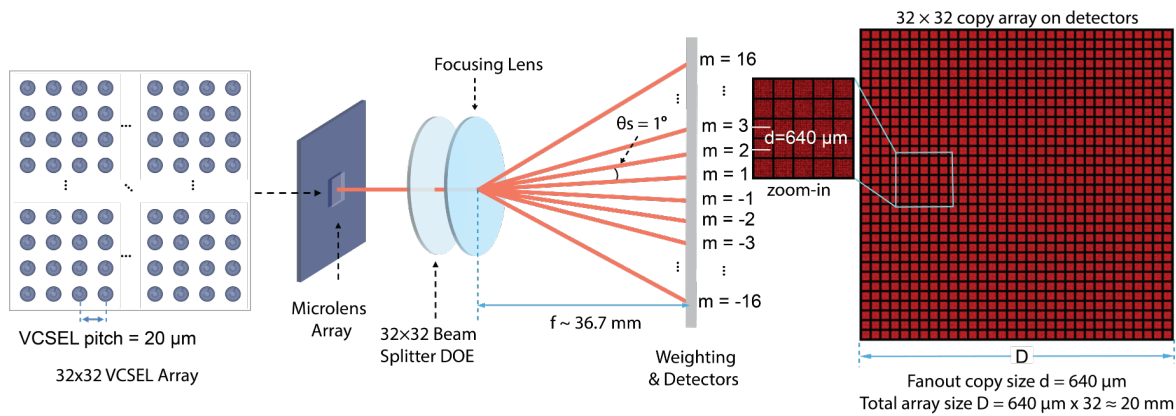

Supplementary Fig. 2. Proposed 32 × 32 VCSEL with 32 × 32 fanout system. A 32 × 32 VCSEL array with 20-µm pitch is collimated by a matching 32 × 32 microlens array, each microlens aligned to a single VCSEL. The beams are replicated into 1024 copies by a 32 × 32 DOE beam splitter and focused onto the weighting/detector plane by a lens. With a diffraction angle of 1 degree per order and a focal length of 36.7 mm, adjacent fanout spots are separated by 640 µm, yielding a 20 × 20 mm$^2$ footprint for the full replica array.

**Aberration control**. We started from a VCSEL array with device pitch of about 10~20 µm, which is available with a micropillar array [16] or a VCSEL array of larger pitch being magnified with a telescope. A 32 × 32 grid at 1 degree per order spans ±16 degrees, which may introduce off-axis aberrations such as coma, astigmatism, field curvature, and distortion. These artifacts induce weighting errors and, in severe cases, crosstalk between adjacent channels. Additional compensation and calibration are therefore required. An aspheric, achromatic objective can be used to suppress field curvature and coma over the full field. For compact builds, a metasurface corrector can supply wide-angle phase compensation with minimal mass[17]. A hybrid refractive-diffractive element in the relay further trims residual off-axis error [18]. Another strategy is calibration and pre-compensation. Incorporate aberration maps into the weighting stage. Apply SLM-based pre-compensation to linearize the field across fanout copies, and use camera-side flat-field gain, a distortion map, and per-spot centroid correction to equalize response [19].

## Signal-to-noise analysis

The basic limitation of optical energy consumption depends on the optical power required to achieve the desired signal-to-noise ratio (SNR) during the detection process, which determines the number of bits of calculation accuracy. In this section, we will model the signal and noise sources in the system and discuss future improvements in computational accuracy and energy consumption.

To establish a quantitative framework, we analyze the performance of each sample using the SNR under equivalent noise bandwidth (ENBW). We denote B as the noise bandwidth of the readout chain, In practice, B can be obtained from $B = 1/(2T) = R/2$, where T is the acquisition time and R is the sampling clockrate. In our intensity detection architecture, the per-sampling signal is represented by $S = P$. Here, $P$ denotes the optical power incident on the photodetector after propagation and SLM modulation during one sample, averaged over the acquisition interval T. Here, we assume the noise comprise three independent terms whose root-mean-square (RMS) input-equivalent power comes from the integration of its power spectral density over B:

$$\text{detector noise: } N_{det} = NEP\sqrt{B},$$

$$\text{shot noise: } N_{shot} = \sqrt{\frac{2h\nu PB}{\eta}},$$

$$\text{source relative-intensity noise (RIN): } N_{RIN} = P\sqrt{(RIN)B}.$$

Where NEP is the noise equivalent power of the photo-receiver. $h$ is Planck's constant, $\nu$ is the frequency, and $\eta$ is the quantum efficiency of the photodetector. RIN is the relative laser intensity noise. Using these definitions, we introduce the per-noise SNRs:

$$SNR_{det} = \frac{P}{N_{det}} = \frac{P\sqrt{2T}}{NEP}, \tag{1}$$

$$SNR_{shot} = \frac{P}{N_{shot}} = \sqrt{\frac{\eta TP}{h\nu}}, \tag{2}$$

$$SNR_{RIN} = \frac{P}{N_{RIN}} = \sqrt{\frac{2T}{RIN}}, \tag{3}$$

The total SNR is

$$\frac{1}{SNR^2} = \frac{1}{SNR_{det}^2} + \frac{1}{SNR_{shot}^2} + \frac{1}{SNR_{RIN}^2}, \tag{4}$$

Leading to

$$SNR = \frac{1}{\frac{1}{SNR_{det}^2} + \frac{1}{SNR_{shot}^2} + \frac{1}{SNR_{RIN}^2}}$$

$$= \frac{P}{\sqrt{(N_{det})^2 + (N_{shot})^2 + (N_{RIN})^2}}$$

$$= \frac{P}{\sqrt{(NEP/\sqrt{2T})^2 + \frac{h\nu P}{\eta T} + (P\sqrt{\frac{RIN}{2T}})^2}} \quad (5)$$

Equation (5) indicates that when at low incident power, $SNR \approx SNR_{det} = \frac{S}{N_{det}} \propto P$, showing detector noise dominates SNR; at high power, $SNR \approx SNR_{RIN} = \frac{S}{N_{SNR}} = SNR_{max}$, showing RIN dominates the SNR and SNR saturates to the RIN plateau. For visualization, we plot the overall SNR as well as the SNR associated with each source individually, as shown in Fig. S3.

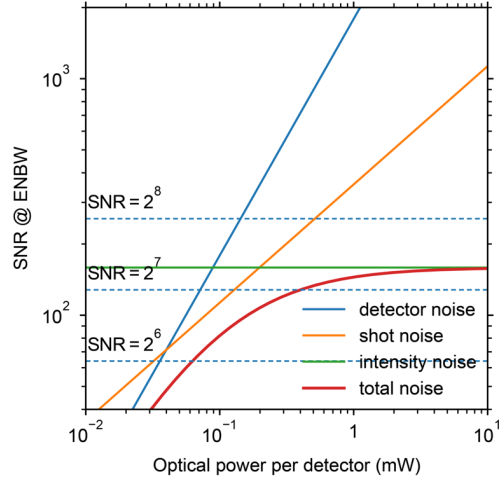

Supplementary Fig. 3. SNR versus optical power per detector. Detector noise, shot noise, and RIN noise are shown together with total SNR. Dashed lines indicate the 6 to 8 bit-precision thresholds. Parameters: R = 25 GS/s (B = 12.5 GHz), $\eta$=0.65, $\lambda$=975 nm, NEP = 5 pW/$\sqrt{HZ}$, RIN = -145dBc/Hz. Noting that this calculation is the numerical accuracy. Adding the sign can increase another computing bit.

The operating point at P=1 mW per detector yields a total SNR≈145 (≈7.2 effective bits), lying above the detector/shot crossovers and close to the RIN-limited plateau. Thus, ~1 mW per detector is sufficient to meet the 7–8-bit precision at 25 GS/s and is consistent with practical per-VCSEL output levels. Noting that in our system the positive and negative signs result in an additional computing bit.